\documentclass[aps,prl,twocolumn,superscriptaddress,english]{revtex4-2}
\usepackage[utf8]{inputenc}

\usepackage{amsmath}
\usepackage{amssymb}
\usepackage{booktabs}
\usepackage{graphicx}
\usepackage{xcolor}
\usepackage{lipsum}
\usepackage{lineno}

\usepackage[version=4]{mhchem}
\usepackage{chemformula}
\usepackage{isotope}

\usepackage{amssymb}
\usepackage{amsmath}
\usepackage{mathtools}
\usepackage{physics}

\usepackage{hyperref}

\newcommand{\tc}{$T_\text{c}\;$}

\makeatletter
\def\@fnsymbol#1{\ensuremath{\ifcase#1\or \dagger\or *\or \ddagger\or
   \mathsection\or \mathparagraph\or \|\or **\or \dagger\dagger
   \or \ddagger\ddagger \else\@ctrerr\fi}}
\makeatother

\usepackage{xr-hyper}
\newcommand*{\addFileDependency}[1]{
\typeout{(#1)}
\@addtofilelist{#1}
\IfFileExists{#1}{}{\typeout{No file #1.}}
}\makeatother

\graphicspath{{plots/}}

\begin{document}

\author{Alessio Cucciari}\email{alessio.cucciari@uniroma1.it}
\affiliation{Dipartimento di Fisica, Sapienza - Universit\`a di Roma, 00185 Rome, Italy}
\author{Dionisia Naddeo} \email{naddeo.1749354@studenti.uniroma1.it}
\affiliation{Dipartimento di Fisica, Sapienza - Universit\`a di Roma, 00185 Rome, Italy}
\author{Simone Di Cataldo}\email{simone.dicataldo@uniroma1.it}
\affiliation{Dipartimento di Fisica, Sapienza - Universit\`a di Roma, 00185 Rome, Italy}
\author{Lilia Boeri} \email{lilia.boeri@uniroma1.it}
\affiliation{Dipartimento di Fisica, Sapienza - Universit\`a di Roma, 00185 Rome, Italy}

\title{\ce{NbTi}: a nontrivial puzzle for the conventional theory of superconductivity}


\date{\today}
\begin{abstract}
We present the first \emph{ab-initio} study of superconductivity in \ce{NbTi}, the workhorse for many applications. Despite its apparent simplicity, \ce{NbTi} turns out to be a major challenge for  computational superconductivity. In fact, anharmonic effects are crucial to obtain dynamically stable phonons for the ordered bcc phase, unstable at the harmonic level, and beyond-Morel Anderson effects in the Coulomb interaction reduce the Tc by more than 20 \%. Lattice disorder causes an additional large discrepancy in \tc compared to experiment. Our results imply that a quantitative description of technologically-relevant superconductors requires methodological developments beyond the current standards. 
\end{abstract}

\maketitle



In the last decade \emph{ab-initio} methods based on Density Functional Theory (DFT)  represented an invaluable driving force in superconducting material research, culminating in the discovery of near-ambient temperature superconductivity in high-pressure superhydrides ~\cite{Duan_SciRep_2014_SH, Drozdov_Nat_SH3_2015, Liu_LaH_2017, Drodzov_Nature_2019_LaH, Somayazulu_PRL_2019_LaH, Boeri_PhysRep_2020_review}.
Wannier-based interpolation of electron-phonon matrix elements \cite{Ponce_CPC_2016_EPW},
accurate functionals for the superconducting state \cite{Sanna_PRL_2020_SCDFT_bench}, improved approximations to describe the Coulomb interaction \cite{Davydov_PRB_2020, Pellegrini_Nb_PRB_2023} and quantum lattice effects \cite{Monacelli_JPCM_2021_SSCHA} are just a few of the recent developments 
which have enormously improved the accuracy of DFT-based methods for superconductors.
Recent benchmarks show that both Superconducting Density Functional Theory (SCDFT) and {\em ab-initio} Migdal-\'Eliashberg 
have achieved an accuracy within 10-20\% against the experimental \tc and related properties for most superconductors \cite{Sanna_PRL_2020_SCDFT_bench,Kawamura_PRB_2020}.
%
However, a few notable exceptions have been found, where 
the description of superconductivity is unusually intricate: a striking example
is elemental mercury \cite{Tresca_PRB_2022}.
\\
\indent
In this Letter we will apply state-of-the-art \emph{ab-initio} methods to study 
\ce{NbTi}, a simple alloy which, due to its relatively low cost, malleability, and high critical fields is 
considered the workhorse material for most
superconducting applications \cite{Scanlan_IEEE, Kinoshita_sc4eng_1990}.
Low-temperature superconductors like \ce{NbTi}
currently account for 75\% of the current market share of superconductors; 
however, despite their technological relevance, they have been
relatively little studied using modern computational techniques.
The only recent first-principle study on \ce{NbTi} 
employs semi-qualitative arguments to explain the 
pressure-dependence of \tc \cite{Zhang_NbTi_PRB_2020}. 
\\
\indent
In this paper we show that obtaining a reliable
estimate of \tc for the simplest model of \ce{NbTi},
i.e. the ordered bcc-$\beta$ phase, requires including non-standard
effects like anharmonic lattice dynamics and energy-dependent
Coulomb interactions.
Even so, the calculated \tc is largely overestimated compared to experiments;
using a simple model based on Boltzmann-averaged supercells
we show that the discrepancy can be qualitatively understood in terms of lattice
disorder. Recovering a quantitative agreement would require 
devising realistic models to include lattice disorder,
which go beyond the current standards of  ab-initio theories of superconductivity.

Commercial samples of \ce{NbTi} contain a mixture of several phases ($\alpha$, $\beta$, $\omega$), but it is normally assumed that superconductivity is dominated by the $\beta$-phase \cite{Collings_Springer_2012}, in which \ce{Nb} and \ce{Ti} atoms are randomly distributed on a bcc lattice.
The bcc $\beta$-phase is the only phase stable for low-Ti concentrations $x\in[0.0, 0.4]$
in the Nb$_{1-x}$Ti$_x$ phase diagram
\cite{Atomic_percentage_footnote}. For larger $x$, increasingly larger inclusions of the hexagonal $\alpha$  form until, for $x \ge 0.8$, the $\alpha$ phase becomes stable. \tc remains essentially constant  $\simeq 9$ K  from $x=1$ until around $x = 0.4$, where it drops rapidly reaching 0.3 K in pure Ti.
The maximum of \tc (9.7~K) is achieved at optimal doping $x_{opt}=0.37$, where the critical field $H_{c2}$ is as high as 14 T at 2 K \cite{Scanlan_IEEE, Collings_Springer_2012}. A recent study has shown that samples near optimal concentration retain 
superconductivity  up to a pressure of 261.7 GPa, setting new records of 19.1 K for the $T_c$ and 19 T for the $H_{c_2}$ for transition metal (TM) alloy superconductors  \cite{Guo_NbTi_2019}. 
\\
\indent
In this work, we employed state-of-the-art ab initio calculations based on Density Functional Theory, the Stochastic Self Consistent Harmonic Approximation (SSCHA), and \'Eliashberg theory for the superconducting state \cite{quantumespresso_1, quantumespresso_2, Errea_PRB_sscha_2014, Monacelli_JPCM_2021_SSCHA, Giustino_PRB_2007_epw, Ponce_CPC_2016_EPW} 
to study the electronic, vibrational and superconducting properties of 
the 1:1 ordered bcc-phase ($x=0.5$)  usually assumed to be a realistic model for near-optimal concentrations -- see \cite{QE_details_footnote} and Section I of the Supplementary Material \cite{suppmat} for computational details. 
In this phase \ce{Nb} and \ce{Ti} atoms occupy alternatively the $1a$ and $1b$ Wyckoff positions of a $Pm\overline{3}m$ \ce{CsCl} lattice \cite{Zhang_NbTi_PRB_2020}; the optimized lattice parameter at ambient pressure is 3.27 Å, to be compared with the experimental value of 3.29 Å. 

\begin{figure}[t]
	\includegraphics[width=1.0\columnwidth]{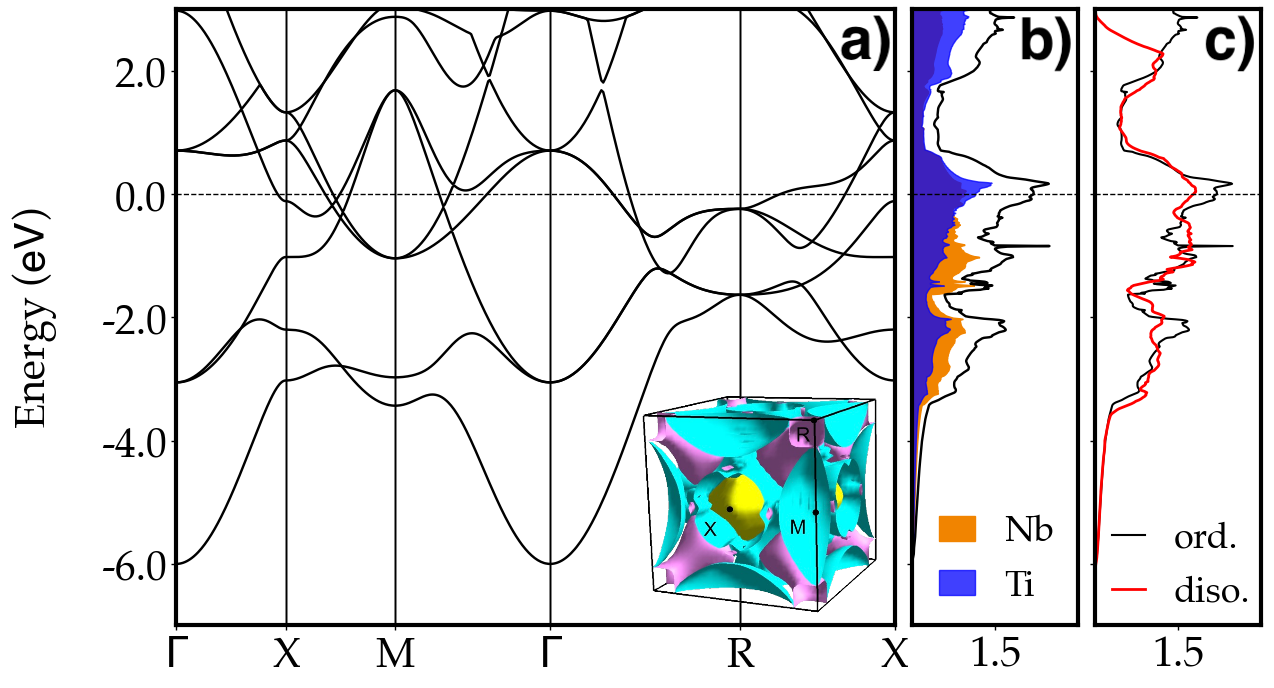}
	\caption{\textbf{a)} Electronic band structure of $\beta$-\ce{NbTi}. The inset shows the Fermi surface, with different colors denoting different bands. \textbf{b)} Atom-projected density of states (DOS) in units of $\text{states/eV/at}$. Projections onto \ce{Nb} and \ce{Ti} are shown in orange and blue, respectively. The zero of the energy is the Fermi level. \textbf{c)} In this panel, the DOS of the ordered phase is compared with that of a simple model of a disordered phase - see text.}
    \label{fig:ebands}
\end{figure}

The electronic structure is shown in Fig.~\ref{fig:ebands}. The valence bands extend $\sim$ 6 eV below the Fermi level, where they are dominated by electronic states of \ce{Nb} and \ce{Ti} \textit{s} and \textit{d}-characters. The corresponding Density of States (DOS), reported in panel b), exhibits several sharp peaks; the Fermi level intersects the DOS at its near-maximum value $N(\epsilon_F)=2.19$ st/eV/at for $x=0.5$. Optimal doping ($x=0.37$)
corresponds to an upward shift of the Fermi level of around 50 meV,
which leaves $N(\epsilon_F)$ almost unchanged \cite{Rigid_band_footnote}. The Fermi surface, shown in the inset of Fig.~\ref{fig:ebands}, comprises four sheets: two large cuboidal hole pockets centered around the $\Gamma$ point and two smaller ellipsoidal pockets 
of electron character around the Brillouin zone edges \cite{FS_decomposed_footnote}. Due to the cuboidal shape of the hole pockets, several portions of the Fermi surface are nearly parallel (nested), causing near-divergences in the non-interacting Lindhard function.

\begin{figure}[b]
	\includegraphics[width=1.05\columnwidth]{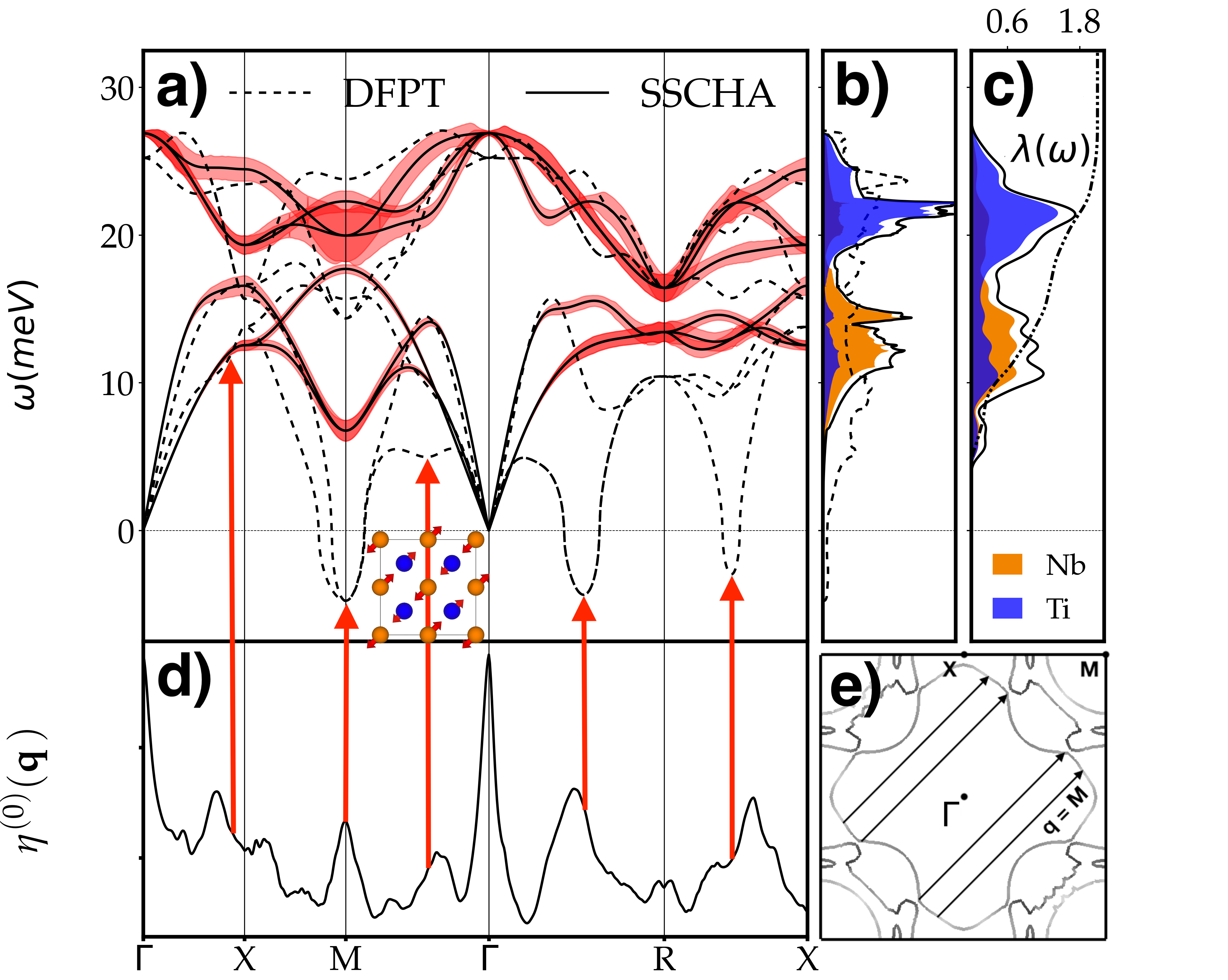}
	\caption{\footnotesize{\textbf{a)} Anharmonic (full black lines) phonon dispersions of the ordered $\beta$-\ce{NbTi}, decorated with their linewidths (red shading). The corresponding harmonic phonons are represented by dashed black lines. The inset shows the vibration pattern of one of the two unstable eigenmodes at the M point. \textbf{b)} Atom-projected and total phonon density of states in units of states/meV. \textbf{c)} Atom-projected and total \'Eliashberg spectral function, along with the electron-phonon coupling $\lambda(\omega)$ (dotdashed black line)
    \textbf{d)} Nesting function $\eta^{(0)}(\mathbf{q})$. Red arrows highlight the position of Kohn anomalies. \textbf{e)} Projection of the Fermi surface onto the [110] lattice plane; black arrows indicate parts of the Fermi surface
    which contribute to the peak in the nesting functions for $\mathbf{q}=M$.
    }}
	\label{fig:sscha}
\end{figure}

The phonon dispersions of ordered $\beta$-\ce{NbTi} are shown in Fig. \ref{fig:sscha} (a). Dashed black lines represent harmonic dispersions obtained within Density Functional Perturbation Theory (DFPT) \cite{Baroni_RevModPhys_2001_DFPT}.
In agreement with previous calculations \cite{Zhang_NbTi_PRB_2020}, we find that at the harmonic level bcc-\ce{NbTi} is dynamically unstable and exhibits imaginary frequencies.
Lattice instabilities, predicted for instance around the M points, midway along the $\Gamma$-R line and the R-X line, are very
localized in momentum space.
This is a typical signature of electronically-driven phonon anomalies -- Kohn anomalies \cite{Kohn_anomalies_PRL_1959}.

In fact, as indicated by the red arrows, the position of the Kohn anomalies in $\mathbf{q}$-space corresponds one-to-one to peaks in the imaginary part of the non-interacting Lindhard function, the so-called nesting function, 
shown in panel $(d)$ of Fig.~\ref{fig:sscha}.
Peaks in the nesting function occur whenever the same wave-vectors $\mathbf{q}$ 
connect, i.e. \emph{nest}  sizable parts of the Fermi surface.
Fig. \ref{fig:sscha} shows for example that the wavevector $\mathbf{q} = \text{M}$ 
is a nesting vector of the Fermi surface, since it connects
nearly-parallel flat faces of the 
hole surface centered at the $\Gamma$ point.

The small inset of panel a) shows the vibration pattern of one of the two modes which are unstable at the harmonic level for this
particular wavevector. The mode induces a shear of the $(0\overline{1}1)$ planes of the bcc crystal, in alternate $[011]$ and $[0\overline{1}\overline{1}]$ directions. This is the Burgers' deformation path for the $\beta \rightarrow \alpha$ (bcc to hexagonal) transition \cite{Burgers_Zr_1934, Djohari_bcc_to_hcp_2009, Grimvall_bcc_to_hcp_2012}, 
which in \ce{Nb_{1-x}Ti_x} is experimentally observed at low temperatures for  $x>0.8$.
For near-optimal concentrations ($x\sim0.5$), the Burgers'
transition is not observed, and the 
bcc phase remains stable at all temperatures \cite{Collings_Springer_2012}. 

Indeed, our calculations show that dynamical instabilities predicted at the harmonic level for $x=0.5$ are eliminated if lattice
dynamics are described with a method which includes
phonon-phonon interactions. In this work we employed a recent implementation  of the Stochastic Self-Consistent Harmonic Approximation which relies on Machine Learning Interatomic Potentials (SSCHA-MLIP \cite{Lucrezi_MLIP_2023}), which ensures an optimal convergence of dynamical matrices and phonon dispersions with respect  both to the number of structures sampled and supercell size. \cite{Details_SSCHA_footnote}.
Solid black lines in Fig. \ref{fig:sscha} (a) represent results of fully anharmonic SSCHA-MLIP calculations \cite{Temp_SSCHA_footnote}: anharmonic effects remove the lattice instabilities at the harmonic level and the residual effect of Kohn anomalies is 
a $\mathbf{q}$-dependent phonon softening.

The sizable anharmonic renormalization of the phonon dispersions in \ce{NbTi} is strongly related to the anharmonic suppression of charge density wave instabilities observed, for instance, in transition metal dichalcogenides \cite{Mauri_PRB_NbSe2_2015}. In both types of systems, sizeable anharmonic effects become relevant not because  atoms with a light mass sample anharmonic regions of the potential corresponding to large displacements, as in hydrides \cite{Mauri_Nature_2016_SH3, Borinaga_PRB_2016_atomicH, Mauri_Nature_2020_LaH}, but because near-divergent electronic  
susceptibilities 
make the phonon potential anharmonic already at small displacements \cite{Boeri_PRB_2002_MgB2_anharm}.

%
To compute the superconducting properties of $\beta$-\ce{NbTi}, we employed the SSCHA dynamical matrices and the electron-phonon matrix elements (calculated from the anharmonic phonon eigenvectors) computed on coarse $\mathbf{k}$- and $\mathbf{q}$-grids in reciprocal space, that were subsequently Wannier-interpolated  on denser grids \cite{Giustino_PRB_2007_epw, Ponce_CPC_2016_EPW} to obtain phonon
frequencies $\omega_{\,\mathbf{q}\nu}$ and linewidths  $\gamma_{\,\mathbf{q}\nu}$, from which the Éliashberg spectral function $\alpha^{2}F(\omega)$ was obtained as:

\begin{equation}
\alpha^{\,2}\,F(\omega) \,= \,\frac{1}{2\pi\,N(\epsilon_F)}\sum_{\;\mathbf{q}\nu}\frac{\gamma_{\mathbf{q}\nu}}{\omega_{\mathbf{q}\nu}}\delta(\omega-\omega_{\mathbf{q}\nu})
\label{alpha_square_iso}
\end{equation}
\vskip2mm

\noindent
The \textit{anharmonic} $\alpha^{2}F(\omega)$, shown in 
panel $c)$ of Fig. \ref{fig:sscha}, closely follows the shape of the phonon DOS, shown in panel $b)$ of the same figure. Indeed,  phonon linewidths, shown as red shading of the phonon dispersions, are distributed uniformly across the whole spectrum.

The phonon spectrum extends up to 28 meV,
with two large peaks centered at 
around 15 meV  and 25 meV, corresponding mainly to  \ce{Nb} and \ce{Ti} vibrations.
It is impossible to compare the calculated
spectrum with experiments, since
no inelastic neutron scattering or 
tunneling data are available for 
\ce{NbTi}. However, both the phonon DOS and the Eliashberg functions resemble closely the data available for pure \ce{Nb}, which has the same bcc crystal structure as NbTi and a very similar \tc of 9.3 K \cite{Wolf_tunnelling_a2F_2011}.

\begin{figure}[t]
	\includegraphics[width=1.0\columnwidth]{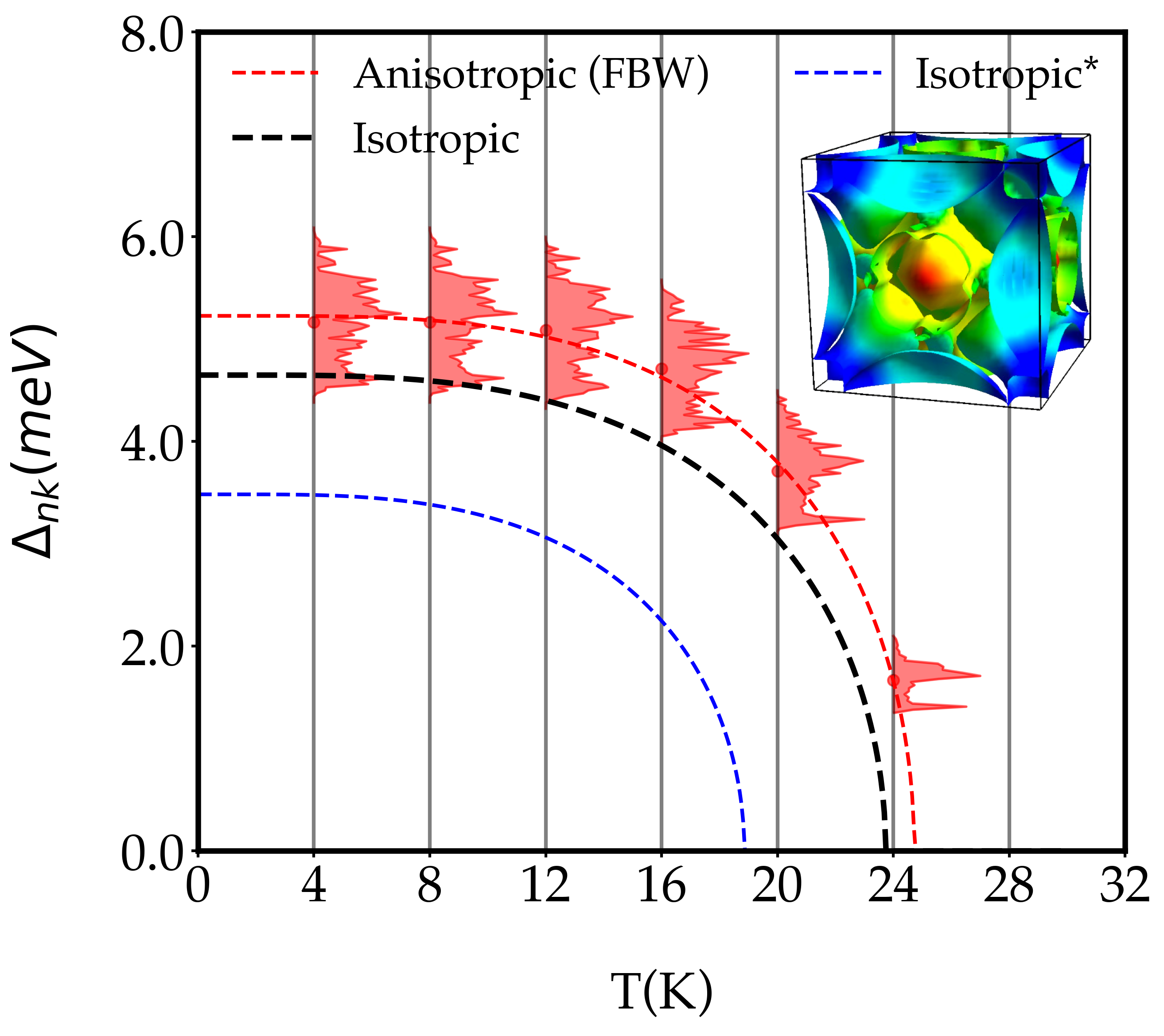}
	\caption{Energy distribution of the zero-frequency superconducting gap of ordered $\beta$-\ce{NbTi} as a function of temperature, obtained solving the anisotropic Migdal–Éliashberg equations for $\mu^*=0.20$. The distribution of the superconducting gap over the Fermi surface is reported in the inset with a color scale which goes from the minimum to the maximum value of the gap (4.5 to 5.8 meV). Dashed lines indicate a fit of the weighted averages of the distribution for each T in red, the corresponding isotropic solution in black, and the isotropic solution for the disordered model in blue.} 
	\label{fig:gap}
\end{figure}

Solving the anisotropic Migdal–Éliashberg equations in the full-bandwidth approximation,
as implemented in the EPW code \cite{Ponce_CPC_2016_EPW, Lucrezi_FBW_2024}, we obtained the superconducting gap as a function of temperature, shown as red dashed line in Fig. \ref{fig:gap}. The  extrapolated value of \tc is 24.3 K \cite{Solving_ME_footnote}, while for the isotropic gap  (black dashed line) we obtain a slightly lower value of \tc (23.7 K).
These values were obtained approximating the residual Coulomb interaction between electrons with a Morel-Anderson pseudopotential $\mu^*=0.20$.

This value, obtained screening the static 
screened interaction computed in the Random-Phase Approximation (details in the SM), is consistent with a similarly large value obtained by Pellegrini et al. for pure \ce{Nb} \cite{Pellegrini_staticCoulomb_2022}.
In the same paper, the authors have shown that, due to the strong localization of $d$ states, treating Coulomb interactions beyond the $\mu^*$ approximation, i.e. in the the full-RPA and beyond-RPA Kukkonen-Overhauser framework, progressively improves the agreement between theoretical and experimental $T_c$'s \cite{Kukkonen_PRB_1979,Pellegrini_Nb_PRB_2023} -- see the second line of table \ref{tab:tc_table}.


Also in \ce{NbTi} using progressively more accurate approximations
for the Coulomb interactions reduces \tc by around 20 $\%$,
down to 19 K in the KO approximation.
However, at variance with \ce{Nb}, in this case the
agreement between calculations and experiments is far
from satisfactory: the calculated \tc
is still twice as large as the experimental one,
 whereas typical errors in $T_c$'s  
for modern \emph{ab-initio} methods for superconductivity are of the order of 10-20 \% \cite{Sanna_PRL_2020_SCDFT_bench,Kawamura_PRB_2020}.
\\
\indent
We can try to retrace the source of this anomalously
large discrepancies by comparing the 
main electron-phonon parameter of \ce{Nb}
and \ce{NbTi},
reported in the first two rows of table \ref{tab:tc_table}. We see that while the values of $\omega_{log}$ are
comparable, the electron-phonon coupling constant $\lambda$ of \ce{NbTi} is 1.5 times as large as in pure \ce{Nb}, reflecting  a similar difference in the value of the DOS at the Fermi level.
Indeed, the high value of $N(\epsilon_F)$
we predict for \ce{NbTi}
may be an artifact of
assuming  a perfectly ordered alloy, whereas in actual samples
\ce{Nb} and \ce{Ti} atoms are randomly distributed on crystal lattice sites. 
The typical effect of disorder
is to “smear out" sharp features
in the electronic DOS and nesting function and hence reduce the magnitude and the concentration-dependence of the $T_c$.
\vskip10pt
\begin{table}[]
\centering
\renewcommand{\arraystretch}{1.5}
\setlength{\tabcolsep}{5pt}
\begin{tabular}{cccccccc}
\hline
\hline
& $N(\epsilon_F)$ & $\omega_{log}$ & $\lambda$ & $T_c^{\,\mu^*}$ & $T_c^{\,\mathrm{RPA}}$ & $T_c^{\,\mathrm{KO}}$ & $T_c^{\,\mathrm{exp}}$ \\
\hline
\ce{NbTi} & 2.19 & 14.4 & 1.87 & 23.7 & 21.4 & 19.0 & 9.7\\
\ce{Nb} & 1.42 & 12.0 & 1.34 & 14.1 & 12.4 & 11.0 & 9.3 \\
\ce{NbTi$^*$} & 1.77 & 14.4 & 1.48 & 18.9 & 16.7 & 14.3 & 9.7 \\
\hline
\hline
\end{tabular}
\caption{Superconducting properties of \ce{Nb} and \ce{NbTi}
alloys. $N(\epsilon_F)$ is the Density of States at the Fermi level in states/ev/at; $\omega_{log}$ is the logarithmically-averaged
phonon frequency in meV; $\lambda$ is the total electron-phonon
coupling parameter. The various \tc in K were obtained solving
the isotropic Migdal-Eliashberg equations using increasingly more accurate approximations for the Coulomb interaction, i.e. 
the Morel-Anderson $\mu^*$ approximation \cite{Morel_PhysRev_1962_mustar}, the full energy-dependent RPA kernel \cite{Pellegrini_staticCoulomb_2022}, and the Kukkonen-Overhauser scheme \cite{Kukkonen_PRB_1979,Pellegrini_Nb_PRB_2023}. Data for pure Nb are from Refs \cite{Sanna_PRL_2020_SCDFT_bench, Sanna_PersonalComm_Nb_2024}. NbTi and NbTi$^*$ indicate results for the ordered bcc lattice and for the disorder-rescaled model (see text).
}
\label{tab:tc_table}
\end{table}

In order to obtain a qualitative estimate of the effect of disorder on the superconducting properties of \ce{NbTi}, we performed a simple \emph{gedanken} experiment, fixing the shape of the Éliashberg function and of the Coulomb interaction to that of the ordered bcc crystal and rescaling their amplitude by the DOS at the Fermi level of a model disordered crystal. 
The DOS of the disordered \ce{NbTi} crystal was simulated performing a Boltzmann-average of the DOS of 100 randomly-generated $2\times2\times2$ supercells with 1:1 stoichiometry \cite{Boltzmann_footnote}. 
The energies of all randomly-generated supercells were in fact down to 60 meV/atom lower than that of
ordered crystal, indicating that 
a randomly disordered crystal may be a more realistic assumption than
the perfectly ordered alloy.
The Boltzmann-averaged DOS is shown as red dashed lines in panel c) of Fig. \ref{fig:ebands}. As expected,  disorder smoothens many of the sharp features of the ordered DOS; as a result the $N(\epsilon_F)$ decreases by 20\%, from 2.19 down to 1.77 st/eV/at. 

Solving the \'Eliashberg equations with the rescaled model,
 we obtain the results shown in the last row of Table \ref{tab:tc_table} (\ce{NbTi$^*$}).
The \tc in the KO approximation is reduced to 14.3 K, in satisfactory agreement with experiment, given 
our crude approximation for disorder.
Some of us have recently proposed that a similar
DOS smoothening effect may explain the weak pressure-dependence of \tc in bcc \ce{Ti} at extreme pressures \cite{Sanna_Ti_PRB_2023}, 
although in that case the underlying physical mechanism 
is based on vacancies rather than disorder.
Both results indicate that the superconducting properties of Ti-rich alloys may be strongly affected by the presence of lattice
imperfections. These types of effects are quite hard to
quantify in first-principles calculations.

In conclusion, in this work we investigated the superconducting
properties of the workhorse superconductor \ce{NbTi},
using state-of-the-art \emph{ab-initio} methods.
We showed that, despite the apparent simplicity of this
material, the description of
the ordered bcc $\beta$-phase, commonly
assumed as a model for the optimally-doped crystal,
presents several challenges for 
state-of-the-art computational methods:
($i$) Lattice dynamics has to be treated in a fully anharmonic framework to remove the dynamical instabilities associated to Kohn anomalies, which prevented the calculations of \tc
in previous studies \cite{Zhang_NbTi_PRB_2020}; ($ii$)
Coulomb interactions beyond the standard Morel-Anderson $\mu^*$
approximation have a non-negligible effect on $T_c$,
reducing the calculated values by more than 20$\%$;
($iii$) This is however not enough to
bring the calculated $T_c$'s in agreement
with experiment, since neglecting
lattice disorder leads to a strong
overestimation of the DOS at the Fermi level, 
and hence $T_c$.
Lattice disorder, which in this work was treated 
with a simple ad-hoc model based on Boltzmann-averaged supercells,
seems to play a major role in many Ti-rich systems,
which  exhibit a weak dependence of \tc 
on external parameters such as composition and pressure \cite{Guo_NbTi_2019, Zhang_NbTi_PRB_2020,Sanna_Ti_PRB_2023}. 
Incorporating qualitatively the effects of lattice
disorder is well beyond the  capabilities
of current first-principles 
theories of superconductivity.
However, this may be 
essential for a real coming-of-age of the field,
as realistic modelling of technologically-relevant
superconductors is arguably one of the essential future challenges
in the field.

\textbf{Acknowledgments}: The authors would like to thank C.Heil, E. Kogler, R. Lucrezi for help with  the SSCHA-MLIP calculations, and A. Sanna for discussion and for performing the calculations of the Coulomb interaction beyond the $\mu^*$ approximation. We would also like to thank Alex Gurevich and David Larbastier for useful discussion on \ce{NbTi} literature.
\\
\indent
The authors acknowledge computational resources from CINECA, projects IsC90-HTS-TECH and IsC99-ACME-C, and the Vienna Scientific Cluster, project 71754 "TEST".
L.B. acknowledges support from
Fondo Ateneo Sapienza 2019-22, and funding from the European Union - NextGenerationEU under the Italian Ministry of University and Research (MUR), “Network 4 Energy Sustainable Transition - NEST” project (MIUR project code PE000021, Concession Degree No. 1561 of October 11, 2022) - CUP C93C22005230007.
\bibliographystyle{apsrev4-1}

\begin{thebibliography}{52}%
	\makeatletter
	\providecommand \@ifxundefined [1]{%
		\@ifx{#1\undefined}
	}%
	\providecommand \@ifnum [1]{%
		\ifnum #1\expandafter \@firstoftwo
		\else \expandafter \@secondoftwo
		\fi
	}%
	\providecommand \@ifx [1]{%
		\ifx #1\expandafter \@firstoftwo
		\else \expandafter \@secondoftwo
		\fi
	}%
	\providecommand \natexlab [1]{#1}%
	\providecommand \enquote  [1]{``#1''}%
	\providecommand \bibnamefont  [1]{#1}%
	\providecommand \bibfnamefont [1]{#1}%
	\providecommand \citenamefont [1]{#1}%
	\providecommand \href@noop [0]{\@secondoftwo}%
	\providecommand \href [0]{\begingroup \@sanitize@url \@href}%
	\providecommand \@href[1]{\@@startlink{#1}\@@href}%
	\providecommand \@@href[1]{\endgroup#1\@@endlink}%
	\providecommand \@sanitize@url [0]{\catcode `\\12\catcode `\$12\catcode
		`\&12\catcode `\#12\catcode `\^12\catcode `\_12\catcode `\%12\relax}%
	\providecommand \@@startlink[1]{}%
	\providecommand \@@endlink[0]{}%
	\providecommand \url  [0]{\begingroup\@sanitize@url \@url }%
	\providecommand \@url [1]{\endgroup\@href {#1}{\urlprefix }}%
	\providecommand \urlprefix  [0]{URL }%
	\providecommand \Eprint [0]{\href }%
	\providecommand \doibase [0]{http://dx.doi.org/}%
	\providecommand \selectlanguage [0]{\@gobble}%
	\providecommand \bibinfo  [0]{\@secondoftwo}%
	\providecommand \bibfield  [0]{\@secondoftwo}%
	\providecommand \translation [1]{[#1]}%
	\providecommand \BibitemOpen [0]{}%
	\providecommand \bibitemStop [0]{}%
	\providecommand \bibitemNoStop [0]{.\EOS\space}%
	\providecommand \EOS [0]{\spacefactor3000\relax}%
	\providecommand \BibitemShut  [1]{\csname bibitem#1\endcsname}%
	\let\auto@bib@innerbib\@empty
	\bibitem [{\citenamefont {Duan}\ \emph {et~al.}(2014)\citenamefont {Duan},
		\citenamefont {Liu}, \citenamefont {Tian}, \citenamefont {Li}, \citenamefont
		{Huang}, \citenamefont {Zhao}, \citenamefont {Yu}, \citenamefont {Liu},
		\citenamefont {Tian},\ and\ \citenamefont {Cui}}]{Duan_SciRep_2014_SH}%
	\BibitemOpen
	\bibfield  {author} {\bibinfo {author} {\bibfnamefont {D.}~\bibnamefont
			{Duan}}, \bibinfo {author} {\bibfnamefont {Y.}~\bibnamefont {Liu}}, \bibinfo
		{author} {\bibfnamefont {F.}~\bibnamefont {Tian}}, \bibinfo {author}
		{\bibfnamefont {D.}~\bibnamefont {Li}}, \bibinfo {author} {\bibfnamefont
			{X.}~\bibnamefont {Huang}}, \bibinfo {author} {\bibfnamefont
			{Z.}~\bibnamefont {Zhao}}, \bibinfo {author} {\bibfnamefont {H.}~\bibnamefont
			{Yu}}, \bibinfo {author} {\bibfnamefont {B.}~\bibnamefont {Liu}}, \bibinfo
		{author} {\bibfnamefont {W.}~\bibnamefont {Tian}}, \ and\ \bibinfo {author}
		{\bibfnamefont {T.}~\bibnamefont {Cui}},\ }\href {\doibase 10.1038/srep06968}
	{\bibfield  {journal} {\bibinfo  {journal} {Scientific Reports}\ }\textbf
		{\bibinfo {volume} {4}},\ \bibinfo {pages} {6968} (\bibinfo {year}
		{2014})}\BibitemShut {NoStop}%
	\bibitem [{\citenamefont {Drozdov}\ \emph {et~al.}(2015)\citenamefont
		{Drozdov}, \citenamefont {Eremets}, \citenamefont {Troyan} \emph
		{et~al.}}]{Drozdov_Nat_SH3_2015}%
	\BibitemOpen
	\bibfield  {author} {\bibinfo {author} {\bibfnamefont {A.}~\bibnamefont
			{Drozdov}}, \bibinfo {author} {\bibfnamefont {M.}~\bibnamefont {Eremets}},
		\bibinfo {author} {\bibfnamefont {I.}~\bibnamefont {Troyan}},  \emph
		{et~al.},\ }\href {\doibase 10.1038/nature14964} {\bibfield  {journal}
		{\bibinfo  {journal} {Nature}\ }\textbf {\bibinfo {volume} {525}},\ \bibinfo
		{pages} {73} (\bibinfo {year} {2015})}\BibitemShut {NoStop}%
	\bibitem [{\citenamefont {Liu}\ \emph {et~al.}(2017)\citenamefont {Liu},
		\citenamefont {Naumov}, \citenamefont {Hoffmann}, \citenamefont {Ashcroft},\
		and\ \citenamefont {Hemley}}]{Liu_LaH_2017}%
	\BibitemOpen
	\bibfield  {author} {\bibinfo {author} {\bibfnamefont {H.}~\bibnamefont
			{Liu}}, \bibinfo {author} {\bibfnamefont {I.~I.}\ \bibnamefont {Naumov}},
		\bibinfo {author} {\bibfnamefont {R.}~\bibnamefont {Hoffmann}}, \bibinfo
		{author} {\bibfnamefont {N.~W.}\ \bibnamefont {Ashcroft}}, \ and\ \bibinfo
		{author} {\bibfnamefont {R.~J.}\ \bibnamefont {Hemley}},\ }\href {\doibase
		10.1073/pnas.1704505114} {\bibfield  {journal} {\bibinfo  {journal}
			{Proceedings of the National Academy of Sciences}\ }\textbf {\bibinfo
			{volume} {114}},\ \bibinfo {pages} {6990} (\bibinfo {year}
		{2017})}\BibitemShut {NoStop}%
	\bibitem [{\citenamefont {Drodzov}\ \emph {et~al.}(2019)\citenamefont
		{Drodzov}, \citenamefont {Kong}, \citenamefont {Besedin}, \citenamefont
		{Kuzonikov}, \citenamefont {Mozaffari}, \citenamefont {Balicas},
		\citenamefont {Balakirev}, \citenamefont {Graf}, \citenamefont {Prakapenka},
		\citenamefont {Greenberg}, \citenamefont {Knyazev}, \citenamefont {Tkacz},\
		and\ \citenamefont {Eremets}}]{Drodzov_Nature_2019_LaH}%
	\BibitemOpen
	\bibfield  {author} {\bibinfo {author} {\bibfnamefont {A.~P.}\ \bibnamefont
			{Drodzov}}, \bibinfo {author} {\bibfnamefont {P.~P.}\ \bibnamefont {Kong}},
		\bibinfo {author} {\bibfnamefont {S.~P.}\ \bibnamefont {Besedin}}, \bibinfo
		{author} {\bibfnamefont {M.~A.}\ \bibnamefont {Kuzonikov}}, \bibinfo {author}
		{\bibfnamefont {S.}~\bibnamefont {Mozaffari}}, \bibinfo {author}
		{\bibfnamefont {L.}~\bibnamefont {Balicas}}, \bibinfo {author} {\bibfnamefont
			{F.~F.}\ \bibnamefont {Balakirev}}, \bibinfo {author} {\bibfnamefont {D.~E.}\
			\bibnamefont {Graf}}, \bibinfo {author} {\bibfnamefont {V.~B.}\ \bibnamefont
			{Prakapenka}}, \bibinfo {author} {\bibfnamefont {E.}~\bibnamefont
			{Greenberg}}, \bibinfo {author} {\bibfnamefont {D.~A.}\ \bibnamefont
			{Knyazev}}, \bibinfo {author} {\bibfnamefont {M.}~\bibnamefont {Tkacz}}, \
		and\ \bibinfo {author} {\bibfnamefont {M.~I.}\ \bibnamefont {Eremets}},\
	}\href {\doibase 10.1038/s41586-019-1201-8} {\bibfield  {journal} {\bibinfo
			{journal} {Nature}\ }\textbf {\bibinfo {volume} {569}},\ \bibinfo {pages}
		{528} (\bibinfo {year} {2019})}\BibitemShut {NoStop}%
	\bibitem [{\citenamefont {Somayazulu}\ \emph {et~al.}(2019)\citenamefont
		{Somayazulu}, \citenamefont {Ahart}, \citenamefont {Mishra}, \citenamefont
		{Geballe}, \citenamefont {Baldini}, \citenamefont {Meng}, \citenamefont
		{Struzhkin},\ and\ \citenamefont {Hemley}}]{Somayazulu_PRL_2019_LaH}%
	\BibitemOpen
	\bibfield  {author} {\bibinfo {author} {\bibfnamefont {M.}~\bibnamefont
			{Somayazulu}}, \bibinfo {author} {\bibfnamefont {M.}~\bibnamefont {Ahart}},
		\bibinfo {author} {\bibfnamefont {A.~K.}\ \bibnamefont {Mishra}}, \bibinfo
		{author} {\bibfnamefont {Z.~M.}\ \bibnamefont {Geballe}}, \bibinfo {author}
		{\bibfnamefont {M.}~\bibnamefont {Baldini}}, \bibinfo {author} {\bibfnamefont
			{Y.}~\bibnamefont {Meng}}, \bibinfo {author} {\bibfnamefont {V.~V.}\
			\bibnamefont {Struzhkin}}, \ and\ \bibinfo {author} {\bibfnamefont {R.~J.}\
			\bibnamefont {Hemley}},\ }\href {\doibase 10.1103/PhysRevLett.122.027001}
	{\bibfield  {journal} {\bibinfo  {journal} {Phys. Rev. Lett.}\ }\textbf
		{\bibinfo {volume} {122}},\ \bibinfo {pages} {027001} (\bibinfo {year}
		{2019})}\BibitemShut {NoStop}%
	\bibitem [{\citenamefont {Flores-Livas}\ \emph {et~al.}(2020)\citenamefont
		{Flores-Livas}, \citenamefont {Boeri}, \citenamefont {Sanna}, \citenamefont
		{Profeta}, \citenamefont {Arita},\ and\ \citenamefont
		{Eremets}}]{Boeri_PhysRep_2020_review}%
	\BibitemOpen
	\bibfield  {author} {\bibinfo {author} {\bibfnamefont {J.~A.}\ \bibnamefont
			{Flores-Livas}}, \bibinfo {author} {\bibfnamefont {L.}~\bibnamefont {Boeri}},
		\bibinfo {author} {\bibfnamefont {A.}~\bibnamefont {Sanna}}, \bibinfo
		{author} {\bibfnamefont {G.}~\bibnamefont {Profeta}}, \bibinfo {author}
		{\bibfnamefont {R.}~\bibnamefont {Arita}}, \ and\ \bibinfo {author}
		{\bibfnamefont {M.}~\bibnamefont {Eremets}},\ }\href {\doibase
		10.1016/j.physrep.2020.02.003} {\bibfield  {journal} {\bibinfo  {journal}
			{Physics Reports}\ }\textbf {\bibinfo {volume} {856}},\ \bibinfo {pages} {1}
		(\bibinfo {year} {2020})}\BibitemShut {NoStop}%
	\bibitem [{\citenamefont {Ponc\'{e}}\ \emph {et~al.}(2016)\citenamefont
		{Ponc\'{e}}, \citenamefont {Margine}, \citenamefont {Verdi},\ and\
		\citenamefont {Giustino}}]{Ponce_CPC_2016_EPW}%
	\BibitemOpen
	\bibfield  {author} {\bibinfo {author} {\bibfnamefont {S.}~\bibnamefont
			{Ponc\'{e}}}, \bibinfo {author} {\bibfnamefont {E.~R.}\ \bibnamefont
			{Margine}}, \bibinfo {author} {\bibfnamefont {C.}~\bibnamefont {Verdi}}, \
		and\ \bibinfo {author} {\bibfnamefont {F.}~\bibnamefont {Giustino}},\ }\href
	{\doibase 10.1016/j.cpc.2016.07.028} {\bibfield  {journal} {\bibinfo
			{journal} {Comp. Phys. Communications}\ }\textbf {\bibinfo {volume} {209}},\
		\bibinfo {pages} {116} (\bibinfo {year} {2016})}\BibitemShut {NoStop}%
	\bibitem [{\citenamefont {Sanna}\ \emph {et~al.}(2020)\citenamefont {Sanna},
		\citenamefont {Pellegrini},\ and\ \citenamefont
		{Gross}}]{Sanna_PRL_2020_SCDFT_bench}%
	\BibitemOpen
	\bibfield  {author} {\bibinfo {author} {\bibfnamefont {A.}~\bibnamefont
			{Sanna}}, \bibinfo {author} {\bibfnamefont {C.}~\bibnamefont {Pellegrini}}, \
		and\ \bibinfo {author} {\bibfnamefont {E.~K.~U.}\ \bibnamefont {Gross}},\
	}\href {\doibase 10.1103/PhysRevLett.125.057001} {\bibfield  {journal}
		{\bibinfo  {journal} {Phys. Rev. Lett.}\ }\textbf {\bibinfo {volume} {125}},\
		\bibinfo {pages} {057001} (\bibinfo {year} {2020})}\BibitemShut {NoStop}%
	\bibitem [{\citenamefont {Davydov}\ \emph {et~al.}(2020)\citenamefont
		{Davydov}, \citenamefont {Sanna}, \citenamefont {Pellegrini}, \citenamefont
		{Dewhurst}, \citenamefont {Sharma},\ and\ \citenamefont
		{Gross}}]{Davydov_PRB_2020}%
	\BibitemOpen
	\bibfield  {author} {\bibinfo {author} {\bibfnamefont {A.}~\bibnamefont
			{Davydov}}, \bibinfo {author} {\bibfnamefont {A.}~\bibnamefont {Sanna}},
		\bibinfo {author} {\bibfnamefont {C.}~\bibnamefont {Pellegrini}}, \bibinfo
		{author} {\bibfnamefont {J.~K.}\ \bibnamefont {Dewhurst}}, \bibinfo {author}
		{\bibfnamefont {S.}~\bibnamefont {Sharma}}, \ and\ \bibinfo {author}
		{\bibfnamefont {E.~K.~U.}\ \bibnamefont {Gross}},\ }\href {\doibase
		10.1103/PhysRevB.102.214508} {\bibfield  {journal} {\bibinfo  {journal}
			{Phys. Rev. B}\ }\textbf {\bibinfo {volume} {102}},\ \bibinfo {pages}
		{214508} (\bibinfo {year} {2020})}\BibitemShut {NoStop}%
	\bibitem [{\citenamefont {Pellegrini}\ \emph {et~al.}(2023)\citenamefont
		{Pellegrini}, \citenamefont {Kukkonen},\ and\ \citenamefont
		{Sanna}}]{Pellegrini_Nb_PRB_2023}%
	\BibitemOpen
	\bibfield  {author} {\bibinfo {author} {\bibfnamefont {C.}~\bibnamefont
			{Pellegrini}}, \bibinfo {author} {\bibfnamefont {C.}~\bibnamefont
			{Kukkonen}}, \ and\ \bibinfo {author} {\bibfnamefont {A.}~\bibnamefont
			{Sanna}},\ }\href {\doibase 10.1103/PhysRevB.108.064511} {\bibfield
		{journal} {\bibinfo  {journal} {Phys. Rev. B}\ }\textbf {\bibinfo {volume}
			{108}},\ \bibinfo {pages} {064511} (\bibinfo {year} {2023})}\BibitemShut
	{NoStop}%
	\bibitem [{\citenamefont {Monacelli}\ \emph {et~al.}(2021)\citenamefont
		{Monacelli}, \citenamefont {Bianco}, \citenamefont {Cherubini}, \citenamefont
		{Calandra}, \citenamefont {Errea},\ and\ \citenamefont
		{Mauri}}]{Monacelli_JPCM_2021_SSCHA}%
	\BibitemOpen
	\bibfield  {author} {\bibinfo {author} {\bibfnamefont {L.}~\bibnamefont
			{Monacelli}}, \bibinfo {author} {\bibfnamefont {R.}~\bibnamefont {Bianco}},
		\bibinfo {author} {\bibfnamefont {M.}~\bibnamefont {Cherubini}}, \bibinfo
		{author} {\bibfnamefont {M.}~\bibnamefont {Calandra}}, \bibinfo {author}
		{\bibfnamefont {I.}~\bibnamefont {Errea}}, \ and\ \bibinfo {author}
		{\bibfnamefont {F.}~\bibnamefont {Mauri}},\ }\href {\doibase
		10.1088/1361-648X/ac066b} {\bibfield  {journal} {\bibinfo  {journal} {J.
				Phys. Condens. Matter}\ }\textbf {\bibinfo {volume} {33}},\ \bibinfo {pages}
		{363001} (\bibinfo {year} {2021})}\BibitemShut {NoStop}%
	\bibitem [{\citenamefont {Kawamura}\ \emph {et~al.}(2020)\citenamefont
		{Kawamura}, \citenamefont {Hizume},\ and\ \citenamefont
		{Ozaki}}]{Kawamura_PRB_2020}%
	\BibitemOpen
	\bibfield  {author} {\bibinfo {author} {\bibfnamefont {M.}~\bibnamefont
			{Kawamura}}, \bibinfo {author} {\bibfnamefont {Y.}~\bibnamefont {Hizume}}, \
		and\ \bibinfo {author} {\bibfnamefont {T.}~\bibnamefont {Ozaki}},\ }\href
	{\doibase 10.1103/PhysRevB.101.134511} {\bibfield  {journal} {\bibinfo
			{journal} {Phys. Rev. B}\ }\textbf {\bibinfo {volume} {101}},\ \bibinfo
		{pages} {134511} (\bibinfo {year} {2020})}\BibitemShut {NoStop}%
	\bibitem [{\citenamefont {Tresca}\ \emph {et~al.}(2022)\citenamefont {Tresca},
		\citenamefont {Profeta}, \citenamefont {Marini} \emph
		{et~al.}}]{Tresca_PRB_2022}%
	\BibitemOpen
	\bibfield  {author} {\bibinfo {author} {\bibfnamefont {C.}~\bibnamefont
			{Tresca}}, \bibinfo {author} {\bibfnamefont {G.}~\bibnamefont {Profeta}},
		\bibinfo {author} {\bibfnamefont {G.}~\bibnamefont {Marini}},  \emph
		{et~al.},\ }\href {\doibase 10.1103/PhysRevB.106.L180501} {\bibfield
		{journal} {\bibinfo  {journal} {Phys. Rev. B}\ }\textbf {\bibinfo {volume}
			{106}},\ \bibinfo {pages} {L180501} (\bibinfo {year} {2022})}\BibitemShut
	{NoStop}%
	\bibitem [{\citenamefont {Scanlan}\ \emph {et~al.}(2004)\citenamefont
		{Scanlan}, \citenamefont {Malozemoff},\ and\ \citenamefont
		{Larbalestier}}]{Scanlan_IEEE}%
	\BibitemOpen
	\bibfield  {author} {\bibinfo {author} {\bibfnamefont {R.}~\bibnamefont
			{Scanlan}}, \bibinfo {author} {\bibfnamefont {A.}~\bibnamefont {Malozemoff}},
		\ and\ \bibinfo {author} {\bibfnamefont {D.}~\bibnamefont {Larbalestier}},\
	}\href {\doibase 10.1109/JPROC.2004.833673} {\bibfield  {journal} {\bibinfo
			{journal} {Proceedings of the IEEE}\ }\textbf {\bibinfo {volume} {92}},\
		\bibinfo {pages} {1639} (\bibinfo {year} {2004})}\BibitemShut {NoStop}%
	\bibitem [{\citenamefont {Kinoshita}(1990)}]{Kinoshita_sc4eng_1990}%
	\BibitemOpen
	\bibfield  {author} {\bibinfo {author} {\bibfnamefont {K.}~\bibnamefont
			{Kinoshita}},\ }\href {\doibase 10.1080/01411599008241820} {\bibfield
		{journal} {\bibinfo  {journal} {Phase Transitions}\ }\textbf {\bibinfo
			{volume} {23}},\ \bibinfo {pages} {73} (\bibinfo {year} {1990})}\BibitemShut
	{NoStop}%
	\bibitem [{\citenamefont {Zhang}\ \emph {et~al.}(2020)\citenamefont {Zhang},
		\citenamefont {Gao}, \citenamefont {Liu},\ and\ \citenamefont
		{Lu}}]{Zhang_NbTi_PRB_2020}%
	\BibitemOpen
	\bibfield  {author} {\bibinfo {author} {\bibfnamefont {J.-F.}\ \bibnamefont
			{Zhang}}, \bibinfo {author} {\bibfnamefont {M.}~\bibnamefont {Gao}}, \bibinfo
		{author} {\bibfnamefont {K.}~\bibnamefont {Liu}}, \ and\ \bibinfo {author}
		{\bibfnamefont {Z.-Y.}\ \bibnamefont {Lu}},\ }\href {\doibase
		10.1103/PhysRevB.102.195140} {\bibfield  {journal} {\bibinfo  {journal}
			{Phys. Rev. B}\ }\textbf {\bibinfo {volume} {102}},\ \bibinfo {pages}
		{195140} (\bibinfo {year} {2020})}\BibitemShut {NoStop}%
	\bibitem [{\citenamefont {Collings}(2012)}]{Collings_Springer_2012}%
	\BibitemOpen
	\bibfield  {author} {\bibinfo {author} {\bibfnamefont {E.}~\bibnamefont
			{Collings}},\ }\href {\doibase 10.1007/978-1-4613-3703-4} {\emph {\bibinfo
			{title} {A sourcebook of titanium alloy superconductivity}}}\ (\bibinfo
	{publisher} {Springer Science \& Business Media},\ \bibinfo {year}
	{2012})\BibitemShut {NoStop}%
	\bibitem [{Ato()}]{Atomic_percentage_footnote}%
	\BibitemOpen
	\href@noop {} {}\bibinfo {note} {In this work alloy compositions are
		specified in atomic percent, rather than weight percent. Thus, $x$ indicates
		the fraction of \ce{Ti} atoms/f.u.}\BibitemShut {Stop}%
	\bibitem [{\citenamefont {Guo}\ \emph {et~al.}(2019)\citenamefont {Guo},
		\citenamefont {Lin}, \citenamefont {Cai}, \citenamefont {Xi}, \citenamefont
		{Zhang}, \citenamefont {Sun}, \citenamefont {Wang}, \citenamefont {Yang},
		\citenamefont {Li}, \citenamefont {Wu}, \citenamefont {Zhang}, \citenamefont
		{Xiang}, \citenamefont {Cava},\ and\ \citenamefont {Sun}}]{Guo_NbTi_2019}%
	\BibitemOpen
	\bibfield  {author} {\bibinfo {author} {\bibfnamefont {J.}~\bibnamefont
			{Guo}}, \bibinfo {author} {\bibfnamefont {G.}~\bibnamefont {Lin}}, \bibinfo
		{author} {\bibfnamefont {S.}~\bibnamefont {Cai}}, \bibinfo {author}
		{\bibfnamefont {C.}~\bibnamefont {Xi}}, \bibinfo {author} {\bibfnamefont
			{C.}~\bibnamefont {Zhang}}, \bibinfo {author} {\bibfnamefont
			{W.}~\bibnamefont {Sun}}, \bibinfo {author} {\bibfnamefont {Q.}~\bibnamefont
			{Wang}}, \bibinfo {author} {\bibfnamefont {K.}~\bibnamefont {Yang}}, \bibinfo
		{author} {\bibfnamefont {A.}~\bibnamefont {Li}}, \bibinfo {author}
		{\bibfnamefont {Q.}~\bibnamefont {Wu}}, \bibinfo {author} {\bibfnamefont
			{Y.}~\bibnamefont {Zhang}}, \bibinfo {author} {\bibfnamefont
			{T.}~\bibnamefont {Xiang}}, \bibinfo {author} {\bibfnamefont {R.~J.}\
			\bibnamefont {Cava}}, \ and\ \bibinfo {author} {\bibfnamefont
			{L.}~\bibnamefont {Sun}},\ }\href {\doibase
		https://doi.org/10.1002/adma.201807240} {\bibfield  {journal} {\bibinfo
			{journal} {Advanced Materials}\ }\textbf {\bibinfo {volume} {31}},\ \bibinfo
		{pages} {1807240} (\bibinfo {year} {2019})}\BibitemShut {NoStop}%
	\bibitem [{\citenamefont {Giannozzi}\ \emph {et~al.}(2009)\citenamefont
		{Giannozzi}, \citenamefont {Baroni}, \citenamefont {Bonini}, \citenamefont
		{Calandra}, \citenamefont {Car}, \citenamefont {Cavazzoni}, \citenamefont
		{Ceresoli}, \citenamefont {Chiarotti}, \citenamefont {Cococcioni},\ and\
		\citenamefont {Dabo}}]{quantumespresso_1}%
	\BibitemOpen
	\bibfield  {author} {\bibinfo {author} {\bibfnamefont {P.}~\bibnamefont
			{Giannozzi}}, \bibinfo {author} {\bibfnamefont {S.}~\bibnamefont {Baroni}},
		\bibinfo {author} {\bibfnamefont {N.}~\bibnamefont {Bonini}}, \bibinfo
		{author} {\bibfnamefont {M.}~\bibnamefont {Calandra}}, \bibinfo {author}
		{\bibfnamefont {R.}~\bibnamefont {Car}}, \bibinfo {author} {\bibfnamefont
			{C.}~\bibnamefont {Cavazzoni}}, \bibinfo {author} {\bibfnamefont
			{D.}~\bibnamefont {Ceresoli}}, \bibinfo {author} {\bibfnamefont {G.~L.}\
			\bibnamefont {Chiarotti}}, \bibinfo {author} {\bibfnamefont {M.}~\bibnamefont
			{Cococcioni}}, \ and\ \bibinfo {author} {\bibfnamefont {I.}~\bibnamefont
			{Dabo}},\ }\href {\doibase 10.1088/0953-8984/21/39/395502} {\bibfield
		{journal} {\bibinfo  {journal} {J. Phys.: Condens. Matter}\ }\textbf
		{\bibinfo {volume} {21}},\ \bibinfo {pages} {395502} (\bibinfo {year}
		{2009})}\BibitemShut {NoStop}%
	\bibitem [{\citenamefont {Giannozzi}\ \emph {et~al.}(2017)\citenamefont
		{Giannozzi}, \citenamefont {Andreussi}, \citenamefont {Brumme}, \citenamefont
		{Bunau}, \citenamefont {Nardelli}, \citenamefont {Calandra}, \citenamefont
		{Car}, \citenamefont {Cavazzoni}, \citenamefont {Ceresoli}, \citenamefont
		{Cococcioni}, \citenamefont {Colonna}, \citenamefont {Carnimeo},
		\citenamefont {Corso}, \citenamefont {de~Gironcoli}, \citenamefont {Delugas},
		\citenamefont {DiStasio}, \citenamefont {Ferretti}, \citenamefont {Floris},
		\citenamefont {Fratesi}, \citenamefont {Fugallo}, \citenamefont {Gebauer},
		\citenamefont {Gerstmann}, \citenamefont {Giustino}, \citenamefont {Gorni},
		\citenamefont {Jia}, \citenamefont {Kawamura}, \citenamefont {Ko},
		\citenamefont {Kokalj}, \citenamefont {K\"{u}c\"{u}kbenli}, \citenamefont
		{Lazzeri}, \citenamefont {Marsili}, \citenamefont {Marzari}, \citenamefont
		{Mauri}, \citenamefont {Nguyen}, \citenamefont {Nguyen}, \citenamefont {de-la
			Roza}, \citenamefont {Paulatto}, \citenamefont {Ponc\'{e}}, \citenamefont
		{Rocca}, \citenamefont {Sabatini}, \citenamefont {Santra}, \citenamefont
		{Schlipf}, \citenamefont {Seitsonen}, \citenamefont {Smogunov}, \citenamefont
		{Timrov}, \citenamefont {Thonhauser}, \citenamefont {Umari}, \citenamefont
		{Vast}, \citenamefont {Wu},\ and\ \citenamefont
		{Baroni}}]{quantumespresso_2}%
	\BibitemOpen
	\bibfield  {author} {\bibinfo {author} {\bibfnamefont {P.}~\bibnamefont
			{Giannozzi}}, \bibinfo {author} {\bibfnamefont {O.}~\bibnamefont
			{Andreussi}}, \bibinfo {author} {\bibfnamefont {T.}~\bibnamefont {Brumme}},
		\bibinfo {author} {\bibfnamefont {O.}~\bibnamefont {Bunau}}, \bibinfo
		{author} {\bibfnamefont {M.~B.}\ \bibnamefont {Nardelli}}, \bibinfo {author}
		{\bibfnamefont {M.}~\bibnamefont {Calandra}}, \bibinfo {author}
		{\bibfnamefont {R.}~\bibnamefont {Car}}, \bibinfo {author} {\bibfnamefont
			{C.}~\bibnamefont {Cavazzoni}}, \bibinfo {author} {\bibfnamefont
			{D.}~\bibnamefont {Ceresoli}}, \bibinfo {author} {\bibfnamefont
			{M.}~\bibnamefont {Cococcioni}}, \bibinfo {author} {\bibfnamefont
			{N.}~\bibnamefont {Colonna}}, \bibinfo {author} {\bibfnamefont
			{I.}~\bibnamefont {Carnimeo}}, \bibinfo {author} {\bibfnamefont {A.~D.}\
			\bibnamefont {Corso}}, \bibinfo {author} {\bibfnamefont {S.}~\bibnamefont
			{de~Gironcoli}}, \bibinfo {author} {\bibfnamefont {P.}~\bibnamefont
			{Delugas}}, \bibinfo {author} {\bibfnamefont {R.~A.}\ \bibnamefont
			{DiStasio}}, \bibinfo {author} {\bibfnamefont {A.}~\bibnamefont {Ferretti}},
		\bibinfo {author} {\bibfnamefont {A.}~\bibnamefont {Floris}}, \bibinfo
		{author} {\bibfnamefont {G.}~\bibnamefont {Fratesi}}, \bibinfo {author}
		{\bibfnamefont {G.}~\bibnamefont {Fugallo}}, \bibinfo {author} {\bibfnamefont
			{R.}~\bibnamefont {Gebauer}}, \bibinfo {author} {\bibfnamefont
			{U.}~\bibnamefont {Gerstmann}}, \bibinfo {author} {\bibfnamefont
			{F.}~\bibnamefont {Giustino}}, \bibinfo {author} {\bibfnamefont
			{T.}~\bibnamefont {Gorni}}, \bibinfo {author} {\bibfnamefont
			{J.}~\bibnamefont {Jia}}, \bibinfo {author} {\bibfnamefont {M.}~\bibnamefont
			{Kawamura}}, \bibinfo {author} {\bibfnamefont {H.-Y.}\ \bibnamefont {Ko}},
		\bibinfo {author} {\bibfnamefont {A.}~\bibnamefont {Kokalj}}, \bibinfo
		{author} {\bibfnamefont {E.}~\bibnamefont {K\"{u}c\"{u}kbenli}}, \bibinfo
		{author} {\bibfnamefont {M.}~\bibnamefont {Lazzeri}}, \bibinfo {author}
		{\bibfnamefont {M.}~\bibnamefont {Marsili}}, \bibinfo {author} {\bibfnamefont
			{N.}~\bibnamefont {Marzari}}, \bibinfo {author} {\bibfnamefont
			{F.}~\bibnamefont {Mauri}}, \bibinfo {author} {\bibfnamefont {N.~L.}\
			\bibnamefont {Nguyen}}, \bibinfo {author} {\bibfnamefont {H.-V.}\
			\bibnamefont {Nguyen}}, \bibinfo {author} {\bibfnamefont {A.~O.}\
			\bibnamefont {de-la Roza}}, \bibinfo {author} {\bibfnamefont
			{L.}~\bibnamefont {Paulatto}}, \bibinfo {author} {\bibfnamefont
			{S.}~\bibnamefont {Ponc\'{e}}}, \bibinfo {author} {\bibfnamefont
			{D.}~\bibnamefont {Rocca}}, \bibinfo {author} {\bibfnamefont
			{R.}~\bibnamefont {Sabatini}}, \bibinfo {author} {\bibfnamefont
			{B.}~\bibnamefont {Santra}}, \bibinfo {author} {\bibfnamefont
			{M.}~\bibnamefont {Schlipf}}, \bibinfo {author} {\bibfnamefont {A.~P.}\
			\bibnamefont {Seitsonen}}, \bibinfo {author} {\bibfnamefont {A.}~\bibnamefont
			{Smogunov}}, \bibinfo {author} {\bibfnamefont {I.}~\bibnamefont {Timrov}},
		\bibinfo {author} {\bibfnamefont {T.}~\bibnamefont {Thonhauser}}, \bibinfo
		{author} {\bibfnamefont {P.}~\bibnamefont {Umari}}, \bibinfo {author}
		{\bibfnamefont {N.}~\bibnamefont {Vast}}, \bibinfo {author} {\bibfnamefont
			{X.}~\bibnamefont {Wu}}, \ and\ \bibinfo {author} {\bibfnamefont
			{S.}~\bibnamefont {Baroni}},\ }\href {\doibase 10.1088/1361-648X/aa8f79}
	{\bibfield  {journal} {\bibinfo  {journal} {J. Phys.: Condens. Matter}\
		}\textbf {\bibinfo {volume} {29}},\ \bibinfo {pages} {465901} (\bibinfo
		{year} {2017})}\BibitemShut {NoStop}%
	\bibitem [{\citenamefont {Errea}\ \emph {et~al.}(2014)\citenamefont {Errea},
		\citenamefont {Calandra},\ and\ \citenamefont
		{Mauri}}]{Errea_PRB_sscha_2014}%
	\BibitemOpen
	\bibfield  {author} {\bibinfo {author} {\bibfnamefont {I.}~\bibnamefont
			{Errea}}, \bibinfo {author} {\bibfnamefont {M.}~\bibnamefont {Calandra}}, \
		and\ \bibinfo {author} {\bibfnamefont {F.}~\bibnamefont {Mauri}},\ }\href
	{\doibase 10.1103/PhysRevB.89.064302} {\bibfield  {journal} {\bibinfo
			{journal} {Phys. Rev. B}\ }\textbf {\bibinfo {volume} {89}},\ \bibinfo
		{pages} {064302} (\bibinfo {year} {2014})}\BibitemShut {NoStop}%
	\bibitem [{\citenamefont {Giustino}\ \emph {et~al.}(2007)\citenamefont
		{Giustino}, \citenamefont {Cohen},\ and\ \citenamefont
		{Louie}}]{Giustino_PRB_2007_epw}%
	\BibitemOpen
	\bibfield  {author} {\bibinfo {author} {\bibfnamefont {F.}~\bibnamefont
			{Giustino}}, \bibinfo {author} {\bibfnamefont {M.~L.}\ \bibnamefont {Cohen}},
		\ and\ \bibinfo {author} {\bibfnamefont {S.~G.}\ \bibnamefont {Louie}},\
	}\href@noop {} {\bibfield  {journal} {\bibinfo  {journal} {Phys. Rev. B}\
		}\textbf {\bibinfo {volume} {76}},\ \bibinfo {pages} {165108} (\bibinfo
		{year} {2007})}\BibitemShut {NoStop}%
	\bibitem [{QE_()}]{QE_details_footnote}%
	\BibitemOpen
	\href@noop {} {}\bibinfo {note} {Electronic and vibrational properties were
		computed within Density Functional Perturbation Theory (DFPT) in a plane-wave
		and pseudopotential framework, as implemented in the Quantum ESPRESSO (QE)
		suite \cite{Baroni_RevModPhys_2001_DFPT, quantumespresso_1}. The wave
		functions expansion was performed with a kinetic energy cutoff of 90.0 Ry. We
		employed the scalar-relativistic version of Optimized Norm-conserving
		Vanderbilt (ONCV) pseudopotentials \cite{Hamann_PRB_2017_ONCV}, with a
		Perdew-Burke-Ernzerhof (PBE) exchange-correlation functional
		\cite{Perdew_PRL_1996_PBE}. The integration over the Brillouin zone was
		carried out using a regular $14\times14\times14$ $\Gamma$-centered
		Monkhorst-Pack \cite{Monkhorst_PRB_1976} grid for electrons, with a
		Methfessel-Paxton smearing of width 0.01 Ry \cite{Methfessel_PRB_1989}, and a
		$6\times6\times6$ mesh for phonons.}\BibitemShut {Stop}%
	\bibitem [{sup()}]{suppmat}%
	\BibitemOpen
	\href@noop {} {}\bibinfo {howpublished} \bibinfo {note} {The
		Supplementary Material [{\url{URL_will_be_inserted_by_publisher}}] contains further computational details on the \emph{ab-initio} calculations, as well as additional figures for the electronic
		structure, comparison between SSCHA-calculated phonon dispersions, details on
		MLIP training,  on the estimate of disorder through the Boltzmann average.} \BibitemShut {Stop}%
	\bibitem [{Rig()}]{Rigid_band_footnote}%
	\BibitemOpen
	\href@noop {} {}\bibinfo {note} {See also Sect. II
		and Fig. 1 of the Supp. Mat.
		\cite{suppmat}}\BibitemShut {NoStop}%
	\bibitem [{FS_()}]{FS_decomposed_footnote}%
	\BibitemOpen
	\href@noop {} {}\bibinfo {note} {A band-by-band decomposition of the Fermi
		surface is reported in Fig. 2 of the
		Supplemental Material.}\BibitemShut {Stop}%
	\bibitem [{\citenamefont {Baroni}\ \emph {et~al.}(2001)\citenamefont {Baroni},
		\citenamefont {de~Gironcoli}, \citenamefont {Corso},\ and\ \citenamefont
		{Giannozzi}}]{Baroni_RevModPhys_2001_DFPT}%
	\BibitemOpen
	\bibfield  {author} {\bibinfo {author} {\bibfnamefont {S.}~\bibnamefont
			{Baroni}}, \bibinfo {author} {\bibfnamefont {S.}~\bibnamefont
			{de~Gironcoli}}, \bibinfo {author} {\bibfnamefont {A.~D.}\ \bibnamefont
			{Corso}}, \ and\ \bibinfo {author} {\bibfnamefont {P.}~\bibnamefont
			{Giannozzi}},\ }\href {\doibase 10.1103/RevModPhys.73.515} {\bibfield
		{journal} {\bibinfo  {journal} {Rev. Mod. Phys}\ }\textbf {\bibinfo {volume}
			{73}},\ \bibinfo {pages} {515} (\bibinfo {year} {2001})}\BibitemShut
	{NoStop}%
	\bibitem [{\citenamefont {Kohn}(1959)}]{Kohn_anomalies_PRL_1959}%
	\BibitemOpen
	\bibfield  {author} {\bibinfo {author} {\bibfnamefont {W.}~\bibnamefont
			{Kohn}},\ }\href {\doibase 10.1103/PhysRevLett.2.393} {\bibfield  {journal}
		{\bibinfo  {journal} {Phys. Rev. Lett.}\ }\textbf {\bibinfo {volume} {2}},\
		\bibinfo {pages} {393} (\bibinfo {year} {1959})}\BibitemShut {NoStop}%
	\bibitem [{\citenamefont {Burgers}(1934)}]{Burgers_Zr_1934}%
	\BibitemOpen
	\bibfield  {author} {\bibinfo {author} {\bibfnamefont {W.}~\bibnamefont
			{Burgers}},\ }\href {\doibase https://doi.org/10.1016/S0031-8914(34)80244-3}
	{\bibfield  {journal} {\bibinfo  {journal} {Physica}\ }\textbf {\bibinfo
			{volume} {1}},\ \bibinfo {pages} {561} (\bibinfo {year} {1934})}\BibitemShut
	{NoStop}%
	\bibitem [{\citenamefont {Djohari}\ \emph {et~al.}(2009)\citenamefont
		{Djohari}, \citenamefont {Milstein},\ and\ \citenamefont
		{Maroudas}}]{Djohari_bcc_to_hcp_2009}%
	\BibitemOpen
	\bibfield  {author} {\bibinfo {author} {\bibfnamefont {H.}~\bibnamefont
			{Djohari}}, \bibinfo {author} {\bibfnamefont {F.}~\bibnamefont {Milstein}}, \
		and\ \bibinfo {author} {\bibfnamefont {D.}~\bibnamefont {Maroudas}},\ }\href
	{\doibase 10.1103/PhysRevB.79.174109} {\bibfield  {journal} {\bibinfo
			{journal} {Phys. Rev. B}\ }\textbf {\bibinfo {volume} {79}},\ \bibinfo
		{pages} {174109} (\bibinfo {year} {2009})}\BibitemShut {NoStop}%
	\bibitem [{\citenamefont {Grimvall}\ \emph {et~al.}(2012)\citenamefont
		{Grimvall}, \citenamefont {Magyari-K\"ope}, \citenamefont {Ozoli\ifmmode
			\mbox{\c{n}}\else \c{n}\fi{}\ifmmode~\check{s}\else \v{s}\fi{}},\ and\
		\citenamefont {Persson}}]{Grimvall_bcc_to_hcp_2012}%
	\BibitemOpen
	\bibfield  {author} {\bibinfo {author} {\bibfnamefont {G.}~\bibnamefont
			{Grimvall}}, \bibinfo {author} {\bibfnamefont {B.}~\bibnamefont
			{Magyari-K\"ope}}, \bibinfo {author} {\bibfnamefont {V.}~\bibnamefont
			{Ozoli\ifmmode \mbox{\c{n}}\else \c{n}\fi{}\ifmmode~\check{s}\else
				\v{s}\fi{}}}, \ and\ \bibinfo {author} {\bibfnamefont {K.~A.}\ \bibnamefont
			{Persson}},\ }\href {\doibase 10.1103/RevModPhys.84.945} {\bibfield
		{journal} {\bibinfo  {journal} {Rev. Mod. Phys.}\ }\textbf {\bibinfo {volume}
			{84}},\ \bibinfo {pages} {945} (\bibinfo {year} {2012})}\BibitemShut
	{NoStop}%
	\bibitem [{\citenamefont {Lucrezi}\ \emph {et~al.}(2023)\citenamefont
		{Lucrezi}, \citenamefont {Kogler}, \citenamefont {Di~Cataldo} \emph
		{et~al.}}]{Lucrezi_MLIP_2023}%
	\BibitemOpen
	\bibfield  {author} {\bibinfo {author} {\bibfnamefont {R.}~\bibnamefont
			{Lucrezi}}, \bibinfo {author} {\bibfnamefont {E.}~\bibnamefont {Kogler}},
		\bibinfo {author} {\bibfnamefont {S.}~\bibnamefont {Di~Cataldo}},  \emph
		{et~al.},\ }\href {\doibase 10.1038/s42005-024-01528-6} {\bibfield  {journal}
		{\bibinfo  {journal} {Communications Physics}\ }\textbf {\bibinfo {volume}
			{6}} (\bibinfo {year} {2023}),\ 10.1038/s42005-024-01528-6}\BibitemShut
	{NoStop}%
	\bibitem [{Det()}]{Details_SSCHA_footnote}%
	\BibitemOpen
	\href@noop {} {}\bibinfo {note} {See also Sect. IV
		and V of the Supplementary Material.}\BibitemShut
	{Stop}%
	\bibitem [{Tem()}]{Temp_SSCHA_footnote}%
	\BibitemOpen
	\href@noop {} {}\bibinfo {note} {The results shown in the figure were
		obtained for T=0 K, but temperature has only a minor effect on the
		dispersions -- See Sect. IV of the Supplementary
		Material.}\BibitemShut {Stop}%
	\bibitem [{\citenamefont {Leroux}\ \emph {et~al.}(2015)\citenamefont {Leroux},
		\citenamefont {Errea}, \citenamefont {Le~Tacon}, \citenamefont {Souliou},
		\citenamefont {Garbarino}, \citenamefont {Cario}, \citenamefont {Bosak},
		\citenamefont {Mauri}, \citenamefont {Calandra},\ and\ \citenamefont
		{Rodi\`ere}}]{Mauri_PRB_NbSe2_2015}%
	\BibitemOpen
	\bibfield  {author} {\bibinfo {author} {\bibfnamefont {M.}~\bibnamefont
			{Leroux}}, \bibinfo {author} {\bibfnamefont {I.}~\bibnamefont {Errea}},
		\bibinfo {author} {\bibfnamefont {M.}~\bibnamefont {Le~Tacon}}, \bibinfo
		{author} {\bibfnamefont {S.-M.}\ \bibnamefont {Souliou}}, \bibinfo {author}
		{\bibfnamefont {G.}~\bibnamefont {Garbarino}}, \bibinfo {author}
		{\bibfnamefont {L.}~\bibnamefont {Cario}}, \bibinfo {author} {\bibfnamefont
			{A.}~\bibnamefont {Bosak}}, \bibinfo {author} {\bibfnamefont
			{F.}~\bibnamefont {Mauri}}, \bibinfo {author} {\bibfnamefont
			{M.}~\bibnamefont {Calandra}}, \ and\ \bibinfo {author} {\bibfnamefont
			{P.}~\bibnamefont {Rodi\`ere}},\ }\href {\doibase 10.1103/PhysRevB.92.140303}
	{\bibfield  {journal} {\bibinfo  {journal} {Phys. Rev. B}\ }\textbf {\bibinfo
			{volume} {92}},\ \bibinfo {pages} {140303} (\bibinfo {year}
		{2015})}\BibitemShut {NoStop}%
	\bibitem [{\citenamefont {Errea}\ \emph {et~al.}(2016)\citenamefont {Errea},
		\citenamefont {Calandra}, \citenamefont {Pickard}, \citenamefont {Nelson},
		\citenamefont {Needs}, \citenamefont {Li}, \citenamefont {Liu}, \citenamefont
		{Zhang}, \citenamefont {Ma},\ and\ \citenamefont
		{Mauri}}]{Mauri_Nature_2016_SH3}%
	\BibitemOpen
	\bibfield  {author} {\bibinfo {author} {\bibfnamefont {I.}~\bibnamefont
			{Errea}}, \bibinfo {author} {\bibfnamefont {M.}~\bibnamefont {Calandra}},
		\bibinfo {author} {\bibfnamefont {C.~J.}\ \bibnamefont {Pickard}}, \bibinfo
		{author} {\bibfnamefont {J.~R.}\ \bibnamefont {Nelson}}, \bibinfo {author}
		{\bibfnamefont {R.~J.}\ \bibnamefont {Needs}}, \bibinfo {author}
		{\bibfnamefont {Y.}~\bibnamefont {Li}}, \bibinfo {author} {\bibfnamefont
			{H.}~\bibnamefont {Liu}}, \bibinfo {author} {\bibfnamefont {Y.}~\bibnamefont
			{Zhang}}, \bibinfo {author} {\bibfnamefont {Y.}~\bibnamefont {Ma}}, \ and\
		\bibinfo {author} {\bibfnamefont {F.}~\bibnamefont {Mauri}},\ }\href
	{\doibase 10.1038/nature17175} {\bibfield  {journal} {\bibinfo  {journal}
			{Nature}\ }\textbf {\bibinfo {volume} {532}},\ \bibinfo {pages} {81}
		(\bibinfo {year} {2016})}\BibitemShut {NoStop}%
	\bibitem [{\citenamefont {Borinaga}\ \emph {et~al.}(2016)\citenamefont
		{Borinaga}, \citenamefont {Errea}, \citenamefont {Calandra}, \citenamefont
		{Mauri},\ and\ \citenamefont {Bergara}}]{Borinaga_PRB_2016_atomicH}%
	\BibitemOpen
	\bibfield  {author} {\bibinfo {author} {\bibfnamefont {M.}~\bibnamefont
			{Borinaga}}, \bibinfo {author} {\bibfnamefont {I.}~\bibnamefont {Errea}},
		\bibinfo {author} {\bibfnamefont {M.}~\bibnamefont {Calandra}}, \bibinfo
		{author} {\bibfnamefont {F.}~\bibnamefont {Mauri}}, \ and\ \bibinfo {author}
		{\bibfnamefont {A.}~\bibnamefont {Bergara}},\ }\href {\doibase
		10.1103/PhysRevB.93.174308} {\bibfield  {journal} {\bibinfo  {journal} {Phys.
				Rev. B}\ }\textbf {\bibinfo {volume} {93}},\ \bibinfo {pages} {174308}
		(\bibinfo {year} {2016})}\BibitemShut {NoStop}%
	\bibitem [{\citenamefont {Errea}\ \emph {et~al.}(2020)\citenamefont {Errea},
		\citenamefont {Belli}, \citenamefont {Monacelli}, \citenamefont {Sanna},
		\citenamefont {Koretsune}, \citenamefont {Tadano}, \citenamefont {Bianco},
		\citenamefont {Calandra}, \citenamefont {Arita}, \citenamefont {Mauri},\ and\
		\citenamefont {Flores-Livas}}]{Mauri_Nature_2020_LaH}%
	\BibitemOpen
	\bibfield  {author} {\bibinfo {author} {\bibfnamefont {I.}~\bibnamefont
			{Errea}}, \bibinfo {author} {\bibfnamefont {F.}~\bibnamefont {Belli}},
		\bibinfo {author} {\bibfnamefont {L.}~\bibnamefont {Monacelli}}, \bibinfo
		{author} {\bibfnamefont {A.}~\bibnamefont {Sanna}}, \bibinfo {author}
		{\bibfnamefont {T.}~\bibnamefont {Koretsune}}, \bibinfo {author}
		{\bibfnamefont {T.}~\bibnamefont {Tadano}}, \bibinfo {author} {\bibfnamefont
			{R.}~\bibnamefont {Bianco}}, \bibinfo {author} {\bibfnamefont
			{M.}~\bibnamefont {Calandra}}, \bibinfo {author} {\bibfnamefont
			{R.}~\bibnamefont {Arita}}, \bibinfo {author} {\bibfnamefont
			{F.}~\bibnamefont {Mauri}}, \ and\ \bibinfo {author} {\bibfnamefont {J.~A.}\
			\bibnamefont {Flores-Livas}},\ }\href {\doibase 10.1038/s41586-020-1955-z}
	{\bibfield  {journal} {\bibinfo  {journal} {Nature}\ }\textbf {\bibinfo
			{volume} {578}},\ \bibinfo {pages} {66} (\bibinfo {year} {2020})}\BibitemShut
	{NoStop}%
	\bibitem [{\citenamefont {Boeri}\ \emph {et~al.}(2002)\citenamefont {Boeri},
		\citenamefont {Bachelet}, \citenamefont {Cappelluti},\ and\ \citenamefont
		{Pietronero}}]{Boeri_PRB_2002_MgB2_anharm}%
	\BibitemOpen
	\bibfield  {author} {\bibinfo {author} {\bibfnamefont {L.}~\bibnamefont
			{Boeri}}, \bibinfo {author} {\bibfnamefont {G.}~\bibnamefont {Bachelet}},
		\bibinfo {author} {\bibfnamefont {E.}~\bibnamefont {Cappelluti}}, \ and\
		\bibinfo {author} {\bibfnamefont {L.}~\bibnamefont {Pietronero}},\ }\href
	{\doibase 10.1103/PhysRevB.65.214501} {\bibfield  {journal} {\bibinfo
			{journal} {Phys. Rev. B}\ }\textbf {\bibinfo {volume} {65}},\ \bibinfo
		{pages} {214501} (\bibinfo {year} {2002})}\BibitemShut {NoStop}%
	\bibitem [{\citenamefont {Wolf}(2011)}]{Wolf_tunnelling_a2F_2011}%
	\BibitemOpen
	\bibfield  {author} {\bibinfo {author} {\bibfnamefont {E.~L.}\ \bibnamefont
			{Wolf}},\ }\href {\doibase 10.1093/acprof:oso/9780199589494.001.0001} {\emph
		{\bibinfo {title} {{Principles of Electron Tunneling Spectroscopy}}}}\
	(\bibinfo  {publisher} {Oxford University Press},\ \bibinfo {year}
	{2011})\BibitemShut {NoStop}%
	\bibitem [{\citenamefont {Lucrezi}\ \emph {et~al.}(2024)\citenamefont
		{Lucrezi}, \citenamefont {Ferreira}, \citenamefont {Hajinazar} \emph
		{et~al.}}]{Lucrezi_FBW_2024}%
	\BibitemOpen
	\bibfield  {author} {\bibinfo {author} {\bibfnamefont {R.}~\bibnamefont
			{Lucrezi}}, \bibinfo {author} {\bibfnamefont {P.~P.}\ \bibnamefont
			{Ferreira}}, \bibinfo {author} {\bibfnamefont {S.}~\bibnamefont {Hajinazar}},
		\emph {et~al.},\ }\href {\doibase 10.1038/s42005-024-01528-6} {\bibfield
		{journal} {\bibinfo  {journal} {Communications Physics}\ }\textbf {\bibinfo
			{volume} {7}} (\bibinfo {year} {2024}),\
		10.1038/s42005-024-01528-6}\BibitemShut {NoStop}%
	\bibitem [{Sol()}]{Solving_ME_footnote}%
	\BibitemOpen
	\href@noop {} {}\bibinfo {note} {See Sect. VII of the
		Supplementary Material for details on the extrapolation.}\BibitemShut {Stop}%
	\bibitem [{\citenamefont {Pellegrini}\ \emph {et~al.}(2022)\citenamefont
		{Pellegrini}, \citenamefont {Heid},\ and\ \citenamefont
		{Sanna}}]{Pellegrini_staticCoulomb_2022}%
	\BibitemOpen
	\bibfield  {author} {\bibinfo {author} {\bibfnamefont {C.}~\bibnamefont
			{Pellegrini}}, \bibinfo {author} {\bibfnamefont {R.}~\bibnamefont {Heid}}, \
		and\ \bibinfo {author} {\bibfnamefont {A.}~\bibnamefont {Sanna}},\ }\href
	{\doibase 10.1088/2515-7639/ac6041} {\bibfield  {journal} {\bibinfo
			{journal} {Journal of Physics: Materials}\ }\textbf {\bibinfo {volume} {5}},\
		\bibinfo {pages} {024007} (\bibinfo {year} {2022})}\BibitemShut {NoStop}%
	\bibitem [{\citenamefont {Kukkonen}\ and\ \citenamefont
		{Overhauser}(1979)}]{Kukkonen_PRB_1979}%
	\BibitemOpen
	\bibfield  {author} {\bibinfo {author} {\bibfnamefont {C.~A.}\ \bibnamefont
			{Kukkonen}}\ and\ \bibinfo {author} {\bibfnamefont {A.~W.}\ \bibnamefont
			{Overhauser}},\ }\href {\doibase 10.1103/PhysRevB.20.550} {\bibfield
		{journal} {\bibinfo  {journal} {Phys. Rev. B}\ }\textbf {\bibinfo {volume}
			{20}},\ \bibinfo {pages} {550} (\bibinfo {year} {1979})}\BibitemShut
	{NoStop}%
	\bibitem [{\citenamefont {Morel}\ and\ \citenamefont
		{Anderson}(1962)}]{Morel_PhysRev_1962_mustar}%
	\BibitemOpen
	\bibfield  {author} {\bibinfo {author} {\bibfnamefont {P.}~\bibnamefont
			{Morel}}\ and\ \bibinfo {author} {\bibfnamefont {P.~W.}\ \bibnamefont
			{Anderson}},\ }\href {\doibase 10.1142/9789812567154_0011} {\bibfield
		{journal} {\bibinfo  {journal} {Physical Review}\ }\textbf {\bibinfo {volume}
			{125}},\ \bibinfo {pages} {1263} (\bibinfo {year} {1962})}\BibitemShut
	{NoStop}%
	\bibitem [{\citenamefont {Pellegrini}\ and\ \citenamefont
		{Sanna}(2024)}]{Sanna_PersonalComm_Nb_2024}%
	\BibitemOpen
	\bibfield  {author} {\bibinfo {author} {\bibfnamefont {C.}~\bibnamefont
			{Pellegrini}}\ and\ \bibinfo {author} {\bibfnamefont {A.}~\bibnamefont
			{Sanna}},\ }\href@noop {} {\bibfield  {journal} {\bibinfo  {journal}
			{Personal Communications}\ } (\bibinfo {year} {2024})}\BibitemShut {NoStop}%
	\bibitem [{Bol()}]{Boltzmann_footnote}%
	\BibitemOpen
	\href@noop {} {}\bibinfo {note} {See Sect. VIII
		and Fig. 8 and 9 of the Supplementary
		Material \cite{suppmat}.}\BibitemShut {Stop}%
	\bibitem [{\citenamefont {Sanna}\ \emph {et~al.}(2023)\citenamefont {Sanna},
		\citenamefont {Pellegrini}, \citenamefont {di~Cataldo}, \citenamefont
		{Profeta},\ and\ \citenamefont {Boeri}}]{Sanna_Ti_PRB_2023}%
	\BibitemOpen
	\bibfield  {author} {\bibinfo {author} {\bibfnamefont {A.}~\bibnamefont
			{Sanna}}, \bibinfo {author} {\bibfnamefont {C.}~\bibnamefont {Pellegrini}},
		\bibinfo {author} {\bibfnamefont {S.}~\bibnamefont {di~Cataldo}}, \bibinfo
		{author} {\bibfnamefont {G.}~\bibnamefont {Profeta}}, \ and\ \bibinfo
		{author} {\bibfnamefont {L.}~\bibnamefont {Boeri}},\ }\href {\doibase
		10.1103/PhysRevB.108.214523} {\bibfield  {journal} {\bibinfo  {journal}
			{Phys. Rev. B}\ }\textbf {\bibinfo {volume} {108}},\ \bibinfo {pages}
		{214523} (\bibinfo {year} {2023})}\BibitemShut {NoStop}%
	\bibitem [{\citenamefont {Hamann}(2017)}]{Hamann_PRB_2017_ONCV}%
	\BibitemOpen
	\bibfield  {author} {\bibinfo {author} {\bibfnamefont {D.~R.}\ \bibnamefont
			{Hamann}},\ }\href {\doibase 10.1103/PhysRevB.88.085117} {\bibfield
		{journal} {\bibinfo  {journal} {Phys. Rev. B}\ }\textbf {\bibinfo {volume}
			{88}},\ \bibinfo {pages} {085117} (\bibinfo {year} {2017})}\BibitemShut
	{NoStop}%
	\bibitem [{\citenamefont {Perdew}\ \emph {et~al.}(1996)\citenamefont {Perdew},
		\citenamefont {Burke},\ and\ \citenamefont
		{Ernzerhof}}]{Perdew_PRL_1996_PBE}%
	\BibitemOpen
	\bibfield  {author} {\bibinfo {author} {\bibfnamefont {J.~P.}\ \bibnamefont
			{Perdew}}, \bibinfo {author} {\bibfnamefont {K.}~\bibnamefont {Burke}}, \
		and\ \bibinfo {author} {\bibfnamefont {M.}~\bibnamefont {Ernzerhof}},\ }\href
	{\doibase 10.1103/PhysRevLett.77.3865} {\bibfield  {journal} {\bibinfo
			{journal} {Phys. Rev. Lett.}\ }\textbf {\bibinfo {volume} {77}},\ \bibinfo
		{pages} {3865} (\bibinfo {year} {1996})}\BibitemShut {NoStop}%
	\bibitem [{\citenamefont {Monkhorst}\ and\ \citenamefont
		{Pack}(1976)}]{Monkhorst_PRB_1976}%
	\BibitemOpen
	\bibfield  {author} {\bibinfo {author} {\bibfnamefont {H.~J.}\ \bibnamefont
			{Monkhorst}}\ and\ \bibinfo {author} {\bibfnamefont {J.~D.}\ \bibnamefont
			{Pack}},\ }\href {\doibase 10.1103/PhysRevB.13.5188} {\bibfield  {journal}
		{\bibinfo  {journal} {Phys. Rev. B}\ }\textbf {\bibinfo {volume} {13}},\
		\bibinfo {pages} {5188} (\bibinfo {year} {1976})}\BibitemShut {NoStop}%
	\bibitem [{\citenamefont {Methfessel}\ and\ \citenamefont
		{Paxton}(1989)}]{Methfessel_PRB_1989}%
	\BibitemOpen
	\bibfield  {author} {\bibinfo {author} {\bibfnamefont {M.}~\bibnamefont
			{Methfessel}}\ and\ \bibinfo {author} {\bibfnamefont {A.~T.}\ \bibnamefont
			{Paxton}},\ }\href {\doibase 10.1103/PhysRevB.40.3616} {\bibfield  {journal}
		{\bibinfo  {journal} {Phys. Rev. B}\ }\textbf {\bibinfo {volume} {40}},\
		\bibinfo {pages} {3616} (\bibinfo {year} {1989})}\BibitemShut {NoStop}%
\end{thebibliography}

\end{document}